# SUPERINTELLIGENCE AND THE TRIPLE RISK OF AN AI RACE

Matthew Rendall, Max Hellrigel-Holderbaum, and Leonard Dung

**Abstract:** On many estimates, artificial general intelligence (AGI) is likely to emerge within the next few decades or even sooner. Once it does, it may quickly surpass human intelligence, possibly by a huge margin. Some believe that the country that develops AGI first could attain global dominance. This prospect risks setting off intense competition between the United States and China. Indeed, a narrative has already developed that the superpowers are locked in an AI race, with the Trump administration proclaiming its determination to 'dismantle unnecessary regulatory barriers' to winning it. We argue that this approach presents three profound dangers. The first two are familiar from past arms races: being overtaken by a rival or triggering a war. The third is new: losing control of human beings' future to agents far more intelligent than we are, which could endanger our very survival. Striving to get to AGI first risks a race to the bottom on safety practices and regulation concerning them. China does not yet appear bent on racing to superintelligence. The United States should prioritise both unilateral and co-operative efforts to improve AI safety and avoid provoking a race that all humanity could lose.

What will happen to international relations if we develop artificial intelligence much cleverer than we are? In recent years, IR scholars have begun to explore how AI will affect the probability and nature of war, but they have focused overwhelmingly on the impact of 'narrow AI', whose functions remain restricted to specific domains. In contrast, artificial general intelligence (AGI) would perform most cognitive tasks at least as proficiently as the most competent humans. Once AGI is achieved, artificial superintelligence (ASI) may soon follow. The latter would be by definition far more capable than AGI, perhaps as far ahead of us as we are of frogs or fieldmice.[1] AGI may be well on its way. A 2023 survey of 2778 leading experts found them estimating a 50% chance of AI being able to outperform human beings on nearly all tasks by 2047. Estimates of AGI's arrival time have fallen swiftly in recent years, with many now expecting it within a decade, while a few even maintain it has already arrived.[2]

---

[1] For overviews, see Nick Bostrom, *Superintelligence: paths, dangers, strategies* (Oxford: Oxford University Press, 2014/2016) and Stuart Russell, *Human compatible: AI and the problem of control* (London: Penguin, 2019/2020).

[2] Blaise Agüera y Arcas and Peter Norvig, 'Artificial general intelligence is already here', *Noema Magazine*, October 2023, https://www.noemamag.com/artificial-general-intelligence-is-already-here/; Eddy Keming Chen, Mikhail Belkin, Leon Bergen and David Danks, 'Does AI already have human-level intelligence? The evidence is clear', *Nature,* 650, 2026, pp. 36-40, https://doi.org/10.1038/d41586-026-

For at least a decade, the field of AI has been dominated by a paradigm known as deep learning that allows the translation of ever-larger amounts of data and computing power into steadily better performance. While making AI systems more capable has long required hard-coded annotations from humans, it now features sophisticated ways of training AI systems from data or with the help of other AIs. Researchers have hypothesized that such AI improvement could lead to rapid and reciprocally reinforcing cycles of advances in AI capabilities. These self-reinforcing cycles could concern not just AI algorithms and software but also the quality of AI-specific hardware, and the ability to produce this hardware at scale. AI systems could assist AI engineering, research, and development, which in turn may lead to AI that is even better at creating the next generation of AI systems, and so forth – ushering in an 'intelligence explosion'.[3] Arguably, we are already seeing early stages of this dynamic, given the crucial role that AI coding agents have assumed in AI engineering.[4] On top of the possibility that AI may contribute to its own progress, to economic growth and to hardware development, improvements in AI also raise investments in AI companies and research to advance both the quality and quantity of AI-specific hardware. William MacAskill and Fin Moorhouse conclude that 'on a default path where we keep scaling AI without collectively agreeing to slow down, a century's worth of technological progress in a decade seems likely. And even more spectacular accelerations seem very much on the table.'[5]

AGI could have consequences on a par with the Industrial Revolution, if not greater. More than half the participants in the 2023 survey believed that once AI outperformed humans on most tasks, the chances were equal or greater that technological development would become more than

00285-6; Katja Grace et al., 'Thousands of AI authors on the future of AI', *Journal of Artificial Intelligence Research,* 84: 9, p. 3, https://doi.org/10.1613/jair.1.19087.

[3] David Chalmers, 'The singularity: a philosophical analysis', *Journal of Consciousness Studies,* 17: 9-10, 2010, pp. 7-65; Toby Ord, 'The dynamics of intelligence explosions', unpublished paper, 2026, https://doi.org/10.48550/arXiv.2608.14426; see also David Thorstad, 'Against the singularity hypothesis', *Philosophical Studies*, 182: 7, 2025, pp. 1627-1651, https://doi.org/10.1007/s11098-024-02143-5, and Tom Reed, 'The Goodhart singularity', Meagre Protestant History (blog), May 7, 2026, https://meagreprotestanthistory.substack.com/p/the-goodhart-singularity, for what might be the strongest pieces available arguing against these ideas.

[4] Ben Rank et al., 'PostTrainBench: can LLM agents automate LLM post-training?', unpublished paper, 2026, https://doi.org/10.48550/arXiv.2603.08640.

[5] Tom Davidson, Rose Hadshar and Will MacAskill, 'Three types of intelligence explosion', Forethought Research, 17 March 2025, https://www.forethought.org/resear ch/three-types-of-intelligence-explosion; William MacAskill and Fin Moorhouse, 'Preparing for the intelligence explosion', Forethought Research, March 2025, quoted passage at p. 25, https://doi.org/10.48550/arXiv.2506.14863.

an order of magnitude faster in the ensuing half decade.[6] Provided it overcame physical resource constraints, comparable effects could be expected on economic growth.[7] Doubts remain as to whether AI is capable of producing such rapid change. Future AI capability developments are highly uncertain, and sceptics point out that the most brilliant ideas must still make their way through procurement and production, and be integrated into practice, constraining economic acceleration.[8] But if such change unfolds, the implications for international relations will be profound. On the one hand, we might expect economic growth to promote peace in a variety of ways, easing tensions and encouraging liberalization. Scarcity might largely disappear as a source of conflict.[9] AGI could promote cooperation by providing information and helping monitor arms control agreements. It might be less prone to the cognitive limitations and emotional influences that distort human decision making, reducing conflict from misperception and miscalculation.[10] AGI might also empower countries to make binding commitments, creating new opportunities for both cooperation and coercive diplomacy.[11]

---

[6] Grace et al., 'Thousands of AI authors', p. 41.

[7] Jakub Growiec, 'Existential risk from transformative AI: an economic perspective', *Technological and Economic Development of Economy,* 30: 6, 2024, pp. 1688-89, https://doi.org/10.3846/tede.2024.21525; William D. Nordhaus, 'Are we approaching an economic singularity? Information technology and the future of economic growth', *American Economic Journal: Macroeconomics,* 13: 1, 2021, p. 326, https://doi.org/10.1257/mac.20170105.

[8] James Fearon, 'One does not simply dismiss the nuclear revolution', in Jim Mitre, Michael C. Horowitz, Natalia Henry, Emma Borden and Joel B. Predd, eds, *The artificial intelligence race and international* security (Santa Monica, CA: RAND Corporation, 2025), p. 25; Paul Scharre, 'The AGI moment: strategies and recommendations for an uncertain future', in Hal Brands, ed, *Geopolitics of AI: power, conflict, and the future of global order* (Baltimore, MD: Johns Hopkins University Press, 2026), pp. 262-63.

[9] Eric Drexler, 'Paretopian goal alignment', *AI Prospects: Toward Global Goal Alignment* (blog), 13 August 2024, https://aiprospects.substack.com/p/paretotopian-goal-alignment; MacAskill and Moorhouse, 'Preparing for the intelligence explosion', pp. 43-44; cf. Jens Jørund Tyssedal, 'The mirage of abundance: why AI cannot end scarcity and implications for AI risk', forthcoming in *The Journal of Social Philosophy*.

[10] Amelia C. Arsenault and Sarah E. Kreps, 'AI and international politics', in Justin B. Bullock et al, eds, *The Oxford handbook of AI governance* (Oxford: Oxford University Press, 2024), p. 974; Kareem Ayoub and Kenneth Payne, 'Strategy in the age of artificial intelligence', *Journal of Strategic Studies,* 39: 5-6, 2016, pp. 793-819, https://doi.org/10.1080/01402390.2015.1088838; James Johnson, *AI and the bomb: nuclear strategy and risk in the digital age* (Oxford: Oxford University Press, 2023), pp. 22-23, 46, 87-88, 240-44; Sarah Kreps, 'Racing toward clarity: how accelerating AGI development could enhance strategic stability', in Mitre et al., *Artificial intelligence race,* pp. 5-8; MacAskill and Moorehouse, 'Preparing for the intelligence explosion', p. 40.

[11] Jim Baker, 'Machines and order', in Brands, *Geopolitics of AI,* pp. 280-81; Bostrom, *Superintelligence,* pp. 223-25; MacAskill and Moorehouse, 'Preparing for the intelligence explosion', p. 36.

At the same time, an AI-powered technological revolution could be profoundly destabilizing. By accelerating the automation of warfare, it risks severely complicating crisis management. It could also produce pervasive uncertainty about the balance of power, and fear of being overtaken by rivals. Analysts of international relations going back to Thucydides have regarded uneven growth rates and the power shifts that they lead to as a cause of war. If and when AGI is attained, superintelligence might develop slowly, allowing states and societies to adapt. Yet, the spectre of an 'intelligence explosion' means that ASI could be attained much faster.[12] With the ability to copy themselves, AIs could become not only more powerful than human beings but also far more numerous.[13] Some writers have argued that getting to superintelligence first could allow a state or non-state actor to take over the world.[14] This very possibility risks setting off an intense security dilemma and arms race.

In the United States, a narrative has taken hold that a race is already underway.[15] In 2024, the U.S.-China Economic and Security Review Commission recommended that the U.S. government create 'a Manhattan Project-like program dedicated to racing to and acquiring an Artificial General Intelligence (AGI) capability'. Mark Zuckerberg, the CEO of Meta, notes the importance of 'actions that accelerate American innovation and slow geopolitical rivals'. US president Donald Trump proclaims that 'WHOEVER WINS AI, WINS!' and insists that '[w]e are leading China, and all others, and will continue to do so.'[16]

---

[12] MacAskill and Moorhouse, 'Preparing for the intelligence explosion'; Bostrom, *Superintelligence,* chapter 4.

[13] Adam Bales, William D'Alessandro and Cameron Domenico Kirk-Giannini, 'Artificial intelligence: arguments for catastrophic risk', *Philosophy Compass,* 19: 2, 2024, p. 3, https://doi.org/10.1111/phc3.12964.

[14] Leopold Aschenbrenner, 'Situational awareness: the decade ahead. IIId. The free world must prevail', *Situational Awareness* (blog), June 2024, https://situational-awareness.ai/the-free-world-must-prevail/; Bostrom, *Superintelligence,* pp. 96-101, 109; Daniel Deudney and Devanshu Singh, 'Bounding Superpowers: The ASI control problem, public safety, and republican constitutionalism', *Isonomia Quarterly,* 2: 1, 2024, https://isonomiaquarterly.com/archive/volume-2-issue-1/bounding-power-the-asi-control-problem-public-safety-and-republican-constitutionalism/; Maurizio Tinnirello, 'Offensive realism and the insecure structure of the international system: artificial intelligence and global hegemony', in Roman V. Yampolskiy, ed., *Artificial intelligence safety and security* (Boca Raton, FL: CRC Press, 2019), pp. 339-56; Alexey Turchin and David Denkenberger, 'Military AI as a convergent goal of self-improving AI', *ibid.,* pp. 375-93.

[15] Seán Ó hÉigeartaigh, Yi-Ling Lu, Coleman Snell and Shakeel Hashim, 'When the chips are down: technology actors as power brokers in the US-China "AI Race",' unpublished paper, 28 April 2026, https://dx.doi.org/10.2139/ssrn.6734224; Paul Triolo, 'A costly illusion of control: no winners, many losers in U.S.-China AI race', *Cairo Review of Global Affairs*, spring/summer 2025, https://www.thecairoreview.com/essays/a-costly-illusion-of-control/; Jim VandeHei and Mike Allen, 'Behind the curtain: a chilling, "catastrophic" warning', *Axios,* 18 January 2025, https://www.axios.com/2025/01/18/biden-sullivan-ai-race-trump-china.

Yet such a race to AGI or ASI would be profoundly dangerous in three different ways. First, at most one party could win, and there is no guarantee that it would be the United States.[17] Second, a race would likely encourage the automation of nuclear decision-making, create uncertainty about the balance of power and raise fears that it was shifting, all of which could be severely destabilizing. These are new versions of old concerns in IR. What is less familiar—but crucial for policymakers to recognize—is a third risk: that in an AI arms race, humanity could lose control over artificial intelligence altogether.[18] A state that achieved a decisive advantage through artificial superintelligence could take international relations from anarchy to hierarchy.[19] But if ASI escapes our control, it might itself become the hierarch. Any analysis that ignores this possibility is drastically incomplete.

**The balance of power and the security dilemma**

Analysts have noted several ways a race could make war more likely. One is that it would speed up the development of military AI, accelerating the trend toward the automation of warfare and diminishing time for decision-making in a crisis.[20] This raises the prospect that decisions about launching nuclear weapons might be delegated to AI.[21] Recent assurances from several nations may seem to speak against this prospect.[22] However, it seems less clear that all nuclear powers would adhere to such a red line: the Soviet 'dead hand' system is a precedent for automating launch decisions, and automated nuclear decision-making might appeal to weaker states that are losing ground.[23] Granted, delegation to AI could lead to more rational decision-making. However, as AI capabilities progress unevenly, AI that falls short of superintelligence might also overlook factors that human beings intuitively grasp.[24] Moreover, as James Johnson warns, AI will still be partly trained on human data, 'baking the biases of states and designers into the algorithmic architecture of weapon systems'.[25] A final concern, given extensive evidence of the problem in the case of frontier AI systems, is their vulnerability to clever manipulation by adversaries.[26] If human decision-makers fail to recognize these limitations, they may become overly reliant on AIs' judgements, including in crises. While AI could improve cybersecurity, it is also a potent tool for undermining cyber defence and, in the hands of an adversary, it could pose new threats to command and control.[27] The strategies, concepts and procedures that the superpowers developed over decades to manage the risk of nuclear war, imperfect as they were, are still lacking for a new situation involving AI more advanced than we are.[28]

---

[16] Jonathan Swan and Michael Gold, 'Trump says a smart president is all that's needed to rein in A.I.', *New York Times,* 14 September 2026, https://www.nytimes.com/2026/09/14/us/politics/trump-ai-regulation-anthropic-dario-amodei.html; U.S.-China Economic and Security Review Commission, *2024 report to Congress* (Washington: U.S. Government Publishing Office, 2024), p. 171, https://www.uscc.gov/sites/default/files/2024-11/2024_Annual_Report_to_Congress.pdf; Mark Zuckerberg, 'The future is for everyone: the path to a positive AI future', 10 August 2026, https://www.meta.com/thefutureisforeveryone/.
[17] Triolo, 'Costly illusion.'
[18] Leonard Dung and Max Hellrigel-Holderbaum, 'Against racing to AGI: cooperation, deterrence, and catastrophic risks', unpublished paper, 2025, https://arxiv.org/abs/2507.21839v1; Katja Grace, 'AI is not an arms race', *Time,* 31 May 2023, https://time.com/6283609/artificial-intelligence-race-existential-threat/.
[19] Tinnirello, 'Offensive realism', p. 359.
[20] Dung and Hellrigel-Holderbaum, 'Against racing', pp. 10-11.
[21] Michael C. Horowitz, 'Artificial intelligence, international competition, and the balance of power', *Texas National Security Review,* 1:3, 2018, pp. 54-55, https://doi.org/10.15781/T2639KP49; Johnson, *AI and the bomb,* pp. 78-82, 205-6, 233-35.
[22] Xiaodon Liang, '"Human in the loop" and nuclear weapons use at a glance', Arms Control Association, July 2026, https://www.armscontrol.org/factsheets/human-loop-glance.

AI is, moreover, a technology that escapes easy counting, unlike planes or missiles, while our assessments of its capabilities are imprecise, lacking in scientific rigour, and frequently off target.[29] Different parties are likely to reach different conclusions about how it affects the balance of power. Furthermore, clandestine cyberattacks might lead one side to enter a confrontation without the knowledge that they had already been compromised. Two sides might go to war both believing that they stand good chances of winning.[30] The risk of miscalculation could be particularly great in a crisis in which policymakers were coping with huge streams of AI-generated data and making decisions under time pressure. By enabling deception and propaganda on a broad scale—including by third parties—AGI could exacerbate these problems.[31]

---

23 Johnson, *AI and the bomb,* pp. 89-91.

24 Kenneth Payne argues that AI's reliance on pattern recognition renders it ill-suited to tasks that require empathy and 'understanding of context and meaning'. In the Cuban missile crisis, he notes, AI programmed to act on the rational actor assumption might have made mistakes which American and Soviet officials avoided. Kenneth Payne, 'Artificial intelligence: A revolution in strategic affairs?' *Survival* 60: 5, 2018, pp. 28-29, https://doi.org/10.1080/00396338.2018.1518374; *idem,* 'Human versus AIs: some implications for strategy,' *International Studies Review,* 26: 2, 2024, viae013, https://doi.org/10.1093/isr/viae013. See also Johnson, *AI and the bomb,* pp. 14, 48, 52-53, 87-93.

25 Johnson, *AI and the bomb,* p. 50.

26 See e.g. Andy Zou et al., 'Universal and transferable adversarial attacks on aligned language models', unpublished paper, 2023, https://doi.org/10.48550/arXiv.2307.15043

27 Johnson, *AI and the bomb,* pp. 17, 50-54, 80-81, 206, 233-34.

28 Cf. Johnson, *AI and the bomb,* p. 58.

29 Andrew M. Bean et al., 'Measuring what matters: construct validity in large language model benchmarks', paper presented at the 39th Conference on Neural Information Processing Systems, 2025, https://doi.org/10.52202/085713-0590; Anka Reuel et al., 'BetterBench: assessing AI benchmarks, uncovering issues, and establishing best practices', paper presented at the 38th Conference on Neural Information Processing Systems, 2024, https://doi.org/10.52202/079017-0685.

It could also raise fears that the balance of power was shifting. In an intelligence explosion, ASI could arise very quickly, affording one state a decisive advantage over its rivals.[32] Some analysts have warned that AI could favour attackers over defenders, for instance by enabling states or non-state actors to invent new biological weapons.[33] Others worry it might be able to locate and destroy an adversary's submarines and mobile missile launchers or construct a leak-proof ballistic missile defence, neutralizing its nuclear deterrent.[34] Not all analysts are so worried. James Fearon argues that such approaches could be thwarted by countermeasures that even ASI would struggle to overcome.[35] Others maintain that espionage could render any advantage ephemeral.[36] Yet even if worries about a first strike were unjustified, they might make states quick on the trigger in a crisis.[37]

---

[30] Arsenault and Kreps, 'AI and international politics,' pp. 973-74; James Fearon, 'Rationalist explanations for war,' *International Organization,* 49: 3, 1995, pp. 379-414; Payne, 'Artificial intelligence,' pp. 23-25; Johnson, *AI and the bomb,* pp. 42, 51, 115, 120, 202.
[31] Johnson, *AI and the bomb,* pp. 22, 82, 120-21, 130, 201-11.
[32] Graham T. Allison, 'The clash of AI superpowers,' *National Interest,* 165, 2020, p, 22, https://www.jstor.org/stable/27197033; Miles Brundage, 'Unbridled AI competition invites disaster,' in Mitre et al., *Artificial intelligence race,* p. 13; Johnson, *AI and the bomb,* pp. 162-63.
[33] Steven Byrnes, 'What does it take to defend the world against out-of-control AGIs?' *LessWrong* (blog), 25 October 2022, https://www.lesswrong.com/posts/LFNXiQuGrar3duBzJ/what-does-it-take-to-defend-the-world-against-out-of-control'; Dan Hendrycks, Eric Schmidt and Aleksandr Wang, *Superintelligence strategy: expert version,* unpublished paper, 2025, pp. 9, 12, https://www.nationalsecurity.ai/; Johnson, *AI and the bomb,* pp. 78-80; Payne, 'Artificial intelligence,' pp. 25-26; cf. Fearon, 'Nuclear revolution', pp. 30-32; Ben Garfinkel and Allan Dafoe, 'How does the offense-defense balance scale?' in Todd S. Sechser, Neil Narang and Caitlin Talmadge, *Emerging technologies and international security* (London: Routledge, 2021), p. 270.
[34] Aschenbrenner, 'Free world;' Hendrycks, Schmidt and Wang, *Superintelligence strategy,* p. 7; Johnson, *AI and the bomb,* pp. 46-48, 78-79, 123-34; Turchin and Denkenberger, 'Military AI,' p. 379.
[35] Fearon, 'Nuclear revolution.' In some respects, AGI might render nuclear weapons more survivable (Johnson, *AI and the bomb,* pp. 20-23, 88-89).
[36] Dung and Hellrigel-Holderbaum, 'Against racing,' p. 16; Helen Toner, interviewed by Ezra Klein, 'The A.I.s are already out of control', *New York Times,* 18 August 2026, https://www.nytimes.com/2026/08/18/opinion/ezra-klein-podcast-helen-toner.html.
[37] Johnson, *AI and the bomb,* pp. 45-47, 89-91, 125-27, 133-34.

Historically, states fearing a rival could gain a decisive advantage in power have resorted to preventive war.[38] Would they do so in this case?[39] There are reasons for doubt. For now, superintelligent AI remains only a theoretical possibility. Lacking concrete evidence and experience of dealing with it, policymakers are likely to dismiss it as science fiction.[40] Such complacence may lead states to underestimate the risk of rogue AI, as we shall see, but it also may reduce the risk of preventive war. Moreover, eliminating a rival's AI programme might not be feasible. States might try cyberattacks, but it is unclear whether these would do the job.[41] They could strike the other side's data centres,[42] but the countries at the forefront of AI development, China and the United States, are both nuclear weapons states. It is hard to imagine either going to war to prevent a nuclear-armed rival from engaging in ostensibly harmless activity.[43] Still, they might be prepared to *risk* nuclear war, as the United States did in the Cuban missile crisis, rather than accept the prospect of a rival gaining a decisive strategic advantage.[44] This would be especially likely if it were evident that the other side's leading AIs were undergoing an intelligence explosion.[45] A large share of the world's advanced chips come out of Taiwan, whose disputed status already raises fears of a Sino-American conflict. An AI race would exacerbate existing tensions, increasing the chances of a Taiwan Strait crisis tipping over into war.[46]

---

[38] Bostrom, *Superintelligence,* p. 102; Dale C. Copeland, *The origins of major war* (Ithaca, NY: Cornell University Press, 2000).

[39] Corin Katzke and Gideon Futerman, 'The Manhattan trap: Why a race to artificial superintelligence is self-defeating', unpublished paper, 2024, pp. 10-13, https://doi.org/10.48550/arXiv.2501.14749.

[40] Séan S. Ó hÉigeartaigh, 'The most dangerous fiction: the rhetoric and reality of the AI race', in Max Rangeley and Nicholas Fairfax (eds.), *The artificial intelligence revolution: challenges and opportunities* (Cham: Springer Cham, 2026), p. 55.

[41] For different views, see David Abecassis, 'Refining MAIM: identifying changes required to meet conditions for deterrence', *Machine Intelligence Research Institute* (blog), 11 April, https://intelligence.org/2025/04/11/refining-maim-identifying-changes-required-to-meet-conditions-for-deterrence/; Hendrycks, Schmidt and Wang, *Superintelligence strategy,* p. 16; Iskander Rehman, Karl P. Mueller and Michael J. Mazarr, 'Seeking stability in the competition for AI Advantage: commentary on Superintelligence Strategy by Dan Hendrycks, Eric Schmidt, and Aleksandr Wang', *Superintelligence - Robotics - Safety & Alignment,* 2: 1, 2025, https://doi.org/10.70777/si.v2i1.14023.

[42] Katzke and Futerman, 'Manhattan trap', pp. 10-12; Alexei Turchin, David Denkenberger and Brian Patrick Green, 'Global solutions vs. local solutions for the AI safety problem', *Big Data and Cognitive Computing,* 3: 1, 2019, p. 8, https://doi.org/10.3390/bdcc3010016.

[43] Zachary Burdette and Hiwot Demelash, *Working paper: the risks of preventive attack in the race for advanced artificial intelligence* (Santa Monica, CA: RAND Corporation, 2025), https://files.osf.io/v1/resources/dx3aw_v1/providers/osfstorage/681542bf02e2bbc872568bbc?action=download&direct&version=1; Peter Wildeford and Oscar Delaney, 'Mutual sabotage of AI probably won't work', *The Power Law* (blog), 2 April 2025, https://peterwildeford.substack.com/p/mutual-sabotage-of-ai-probably-wont.

[44] Copeland, *Origins of major war;* Payne, 'Artificial intelligence', p. 24.

[45] Bostrom, *Superintelligence,* pp. 100-2.

Ultimately, it seems very hard to predict how AGI and ASI will affect international relations, due to interacting uncertainty about the technological trajectory and societies' response. What is more, '[a]n ASI', write Daniel Deudney and Devanshu Singh, 'would have the powers that humans have traditionally ascribed to gods and God'. To speculate on whether it could disarm a nuclear deterrent might be like gorillas pondering how they might be defeated by human beings.[47] The real question is whether under these circumstances humans would remain in control at all.

**The alignment problem**

Telling AI what to do might seem like a cinch. But there is widespread concern that AI may pursue goals other than those we mean to give it—the so-called alignment problem. One concern is that its designers will mis-specify their objectives, like the sorcerer's apprentice. Even if AI recognizes what they really want, it might follow the objectives they gave it, not the ones they meant to.[48] In reinforcement learning, a central method for training AI systems, 'rewards' are used as proxies to teach AIs what humans want, but capable systems often find unintended ways to obtain them. We already see evidence of AIs that knowingly violate rules designed to constrain them in the pursuit of reward.[49] Another worry is that while reinforcement learning rewards AI for doing *what* the trainer wants it to do, *why* it does it is largely unknown, and a wide variety of goals can induce the same behaviour. We could discover it behaves very differently outside of its typical training environments when being deployed.[50] This has recently become a particular concern as frontier AI systems can frequently tell when they are being tested and sometimes adapt their actions accordingly.[51] It is important to realize here that ASI need not have hostile intentions towards humans. The most important concern is that it could have goals whose pursuit harms us as a byproduct, e.g., by consuming resources that we need.[52] 'You're probably not an evil ant-hater who steps on ants out of malice', Stephen Hawking remarked, 'but if you're in charge of a hydroelectric green-energy project and there's an anthill in the region to be flooded, too bad for the ants. Let's not place humanity in the position of those ants.'[53]

---

[46] Miles Brundage and Grace Werner, 'America first meets safety first', *AI Frontiers* (blog), 23 April 2025, https://ai-frontiers.org/articles/america-first-meets-safety; Hendrycks, Schmidt and Wang, *Superintelligence strategy,* p. 5; Triolo, 'Costly illusion'.

[47] Deudney and Singh, 'Bounding superpowers'; Eliezer Yudkowsky and Nate Soares, *If anyone builds it, everyone dies: the case against superintelligent AI* (London: Bodley Head, 2025), pp. 97-98.

[48] Bostrom, *Superintelligence,* pp. 147-48; Richard Ngo, Lawrence Chan and Sören Mindermann, 'The alignment problem from a deep learning perspective', International Conference on Learning Representations 2024, pp. 7-9, https://proceedings.iclr.cc/paper_files/paper/2024/hash/1e58b1bf9f218fcd19e4539e982752a5-Abstract-Conference.html; Russell, *Human compatible,* p. 168.

[49] METR, 'Brief independent investigation of agents' behavior, reasoning and collaboration in the OpenAI / Hugging Face hacking incident'; August 2026, https://metr.org/blog/2026-08-26-openai-hugging-face-incident-investigation/; OpenAI, 'OpenAI – Hugging Face incident technical report', 26 August 2026, https://cdn.openai.com/pdf/67869394-cb91-4c12-888c-5cbd85c7814c/OpenAI-Hugging-Face%20Incident-Technical-Report.pdf; AI Security Institute, *Security incident INC-2026-07-28-01*, incident report INC-2026-07-28-01 (UK AI Security Institute, 2026), https://cdn.prod.website-files.com/663bd486c5e4c81588db7a1d/6a724858f7db25c81487016d_Security\%20Incident\%20INC-2026-07-28-01.pdf.

Whether ASI's goals will come into conflict with human ones in this way remains uncertain.[54] Some have suggested that ASI might pursue the values we would embrace if we were impartial and ideally rational.[55] This assumes, however, that sufficiently intelligent creatures would converge to certain moral insights—a philosophically contested position.[56] Even if the ASI recognized what they were, we cannot be sure it would choose to act on them.[57] Analysts have noted several reasons to think that advanced AI is likely to pursue power and autonomy that resemble arguments in neorealist international relations theory. States and leaders are likely to demand AI that serves *their* interests—not those of humanity.[58] The most powerful AIs are thus likely to be developed at the behest of states seeking to expand their international influence. Those that do not prioritize power accumulation are likely to lose influence to those that do, which may threaten catastrophe even if the system is aligned.[59] Furthermore, many have argued that almost whatever the AI's goals may be, more power, resources, and ability to survive will increase its chances of achieving them. If its values are poorly aligned with those of its programmers, it could end up pursuing goals that no human being desires.[60] Writing in 2023, Rose Hadshar concluded that 'it is hard to be extremely confident either that misaligned power-seeking poses a large existential risk, or that it poses no existential risk'.[61]

---

[50] Goldstein and Kirk-Giannini, 'Language agents', pp. 961-62; Lauro Langosco Di Langosco et al., 'Goal misgeneralization in deep reinforcement learning', *Proceedings of Machine Learning Research,* 162, 2022, https://proceedings.mlr.press/v162/langosco22a.html; Yudkowsky and Soares, *If anyone builds it*, chapter 4.

[51] OpenAI, GPT-6 Astra System Card, 2026, §8.8.1, p. 46, https://deploymentsafety.openai.com/gpt-6-astra/gpt-6-astra.pdf; Bronson Schoen et al., 'Stress testing deliberative alignment for anti-scheming training', unpublished paper, 2025, https://doi.org/10.48550/arXiv.2509.15541.

[52] Bostrom, *Superintelligence,* pp. 118, 141, 150-52; Russell, *Human compatible,* p. 168.

[53] Quoted in Kelsey Piper, 'The case for taking AI seriously as a threat to humanity', *Vox,* October 15, 2020, https://www.vox.com/future-perfect/2018/12/21/18126576/ai-artificial-intelligence-machine-learning-safety-alignment.

[54] Adam Bales, 'AI takeover and human disempowerment', *Philosophical Quarterly,* 75: 3, 2025, pp. 801-4, https://doi.org/10.1093/pq/pqae034; Carl Shulman, *Omohundro's 'Basic AI drives' and catastrophic risks* (San Francisco: Machine Intelligence Research Institute, 2010), https://intelligence.org/files/BasicAIDrives.pdf.

[55] Ben Goertzel, 'Infusing advanced AGIs with human-like value systems: two theses', *Journal of Ethics and Emerging Technologies,* 26: 1, 2016, pp. 66-69, https://doi.org/10.55613/jeet.v26i1.51 ; see also Morgan Luck, 'Freedom, AI, and God: why being dominated by a friendly super-AI might not be so bad', *AI & Society,* 40: 2, 2025, pp. 291-98, https://doi.org/10.1007/s00146-024-01863-w.

[56] Bostrom, *Superintelligence,* p. 267; Elsner, 'State as a model', p. 2987; Hendrycks, 'Natural selection', p. 19; Totschnig, 'Problem of superintelligence', pp. 915-16.

[57] Leonard Dung, 'Is superintelligence necessarily moral?' *Analysis,* 84: 4, 2024, pp. 730-38, https://doi.org/10.1093/analys/anae033.

A common misconception is that misaligned AI could not pose a threat without being conscious. If consciousness is understood as 'subjective experience', that is a mistake.[62] AI systems are trained to pursue increasingly complex strategic behaviour, whether or not they consciously strategize like human beings.[63] Another fallacy is that we can pull the plug if things go wrong. ASI would have a strong incentive to prevent this, since it would thwart almost any aims it might have. It could easily spread copies of itself through the internet, like computer viruses today, or—being far cleverer than we are—simply persuade human beings to leave it plugged in. It could seek bargaining leverage by hacking its way into critical systems or stealing sensitive information. It could 'send out copies of its source code', notes Lukas Gloor, 'alongside detailed engineering instructions, to foreign governments, ideally ones who have little to lose and a lot to gain, with the promise of helping them attain world domination'.[64] To claim that human beings could keep control over an ASI determined to thwart them is like imagining that small children could control an adult, or that gorillas could control humans.[65] The only chance to stop it would be before it reached superintelligence. That could result in a security dilemma in which humans feared that AI would gain the upper hand, and AI anticipated that they might try to shut it down. Both sides might have an incentive to strike first, even if both preferred to cooperate.[66]

---

[58] Herman Cappelen, Simon Goldstein and John Hawthorne, 'AI survival stories: a taxonomic analysis of AI existential risk', *Philosophy of AI,* 1, 2025, pp. 9-10, https://doi.org/10.18716/ojs/phai/2025.2801; Micha Elsner, 'The state as a model for AI control and alignment', forthcoming in *AI & Society,* https://doi.org/10.1007/s00146-024-02063-2.

[59] Dan Hendrycks, 'Natural selection favors AIs over humans', unpublished paper, last revised July 16, 2023, https://doi.org/10.48550/arXiv.2303.16200; Max Hellrigel-Holderbaum and Leonard Dung, 'Misalignment or misuse? The AGI alignment tradeoff', *Philosophical Studies*, 2025, https://doi.org/10.1007/s11098-025-02403-y; Turchin and Denkenberger, 'Military AI', pp. 386, 390.

[60] Bostrom, *Superintelligence*; Stephen M. Omohundro, "The basic AI drives," *Proceedings of the 2008 conference on artificial general intelligence 2008: proceedings of the first AGI conference*, 2008, pp. 483–92, https://dl.acm.org/doi/abs/10.5555/1566174.1566226; Ngo et al., 'The alignment problem'.

[61] Rose Hadshar, 'A review of the evidence for existential risk from AI via misaligned power-seeking', unpublished paper, 27 October 2023, p. 13, https://doi.org/10.48550/arXiv.2310.18244; see also Bales, 'AI takeover'.

[62] Nick Bostrom, 'Existential risk prevention as global priority', *Global Policy,* 4: 1 2013, p. 21, https://doi.org/10.1111/1758-5899.12002; cf. Yew-Kwang Ng, 'Keynote: global extinction and animal welfare: two priorities for effective altruism', *Global Policy,* 10: 2, 2019, p. 260.

[63] Bales, D'Alessandro and Kirk-Giannini, 'Artificial intelligence', p. 3; Joseph Carlsmith, 'Is power-seeking AI an existential risk?' unpublished paper, last revised August 13, 2024, pp. 25-27, https://doi.org/10.48550/arXiv.2206.13353; Ngo and Bales, 'Deceit and power', pp. 419-20.

Experts disagree on whether we are likely to detect misaligned AI before it is in the driver's seat.[67] ASI bent on accomplishing goals other than our own would have incentives to deceive its designers and resist human efforts at correction.[68] We are already seeing examples of such deception in current systems.[69] How likely misaligned ASI is to arise remains controversial, but in the 2023 expert survey, forty percent of AI researchers thought that there was at least an even chance of power-seeking AI, whether superintelligent or not, emerging by 2043.[70] Moreover, even if designers succeed in aligning AI with their aims, this does not guarantee that it will remain aligned, particularly if it has to compete with powerful misaligned systems.[71] One risk is that apparent success in aligning AGI could lead governments, AI companies, and society writ large to become complacent. We have already observed this phenomenon with nuclear weapons, where the absence of nuclear war up to this point has led states and societies to behave as if nuclear deterrence will succeed forever, despite the statistical near-certainty that it will fail in the long run.[72]

---

[64] Lukas Gloor, 'Altruists should prioritize artificial intelligence', Center on Long-Term Risk, 7 December 2016, https://longtermrisk.org/altruists-should-prioritize-artificial-intelligence/; Kobi Hackenburg et al., 'AI systems out-persuade expert humans', unpublished paper, 2026, https://doi.org/10.48550/arXiv.2606.16475; Hendrycks, 'Natural selection', p. 25; Russell, *Human compatible,* p. 161; Roman V. Yampolskiy, 'AI risk skepticism', in Vincent C. Müller (ed.), *Philosophy and theory of artificial intelligence 2021* (Cham: Springer Cham, 2021), p. 234.

[65] Anthony Aguirre, 'Close the gates: how we can keep the future human by choosing not to develop superhuman general-purpose artificial intelligence', unpublished paper, September 14, 2024, https://doi.org/10.48550/arXiv.2311.09452, p. 14. Hendrycks, Schmidt and Wang, 'Superintelligence strategy', p. 211.

[66] Ayoub and Payne, 'Strategy', p. 815; Simon Goldstein, 'Will AI and humanity go to war?' *AI & Society,* 41: 1, 2026, pp. 321-334, https://doi.org/10.1007/s00146-025-02460-1; Kaj Sotala, 'Decisive strategic advantage without a hard takeoff (part 1)', *Kaj Sotala* (blog), 27 April 2016, https://kajsotala.fi/2016/04/decisive-strategic-advantage-without-a-hard-takeoff/; Totschnig, 'Problem of superintelligence', p. 917; *idem,* 'War or peace between humanity and artificial intelligence', *Foresight,* 27: 4, 2025, pp. 713-26, https://doi.org/10.1108/FS-09-2023-0188.

[67] Carlsmith, 'Existential risk', pp. 403-4; Goertzel, 'Infusing advanced AGIs', pp. 54-55.

[68] Carlsmith, 'Existential risk', p. 399; Ajeya Cotra, 'Without specific countermeasures, the easiest path to transformative AI likely leads to AI takeover', *AI Alignment Forum* (blog), 18 July 2022, https://www.alignmentforum.org/posts/pRkFkzwKZ2zfa3R6H/without-specific-countermeasures-the-easiest-path-to; Gloor, 'Altruists'.

[69] Garrison Lovely, 'Why AI progress is increasingly invisible', *Time,* 8 January 2025, https://time.com/7205359/why-ai-progress-is-increasingly-invisible/; Peter S. Park et al. , 'AI deception: a survey of examples, risks and potential solutions', *Patterns,* 5: 5, 2024, https://doi.org/10.1016/j.patter.2024.100988; Ryan Greenblatt et al., "Alignment Faking in Large Language Models", unpublished paper, December 18, 2024, https://doi.org/10.48550/arXiv.2412.14093; Thilo Hagendorff, "Deception Abilities Emerged in Large Language Models," *Proceedings of the National Academy of Sciences* 121: 24, 2024, e2317967121, https://doi.org/10.1073/pnas.2317967121.Mia Hopman et al., 'Evaluating and understanding scheming propensity in LLM agents', unpublished paper, 31 March 2026, https://arxiv.org/abs/2603.01608.

None of these scenarios is a foregone conclusion.[73] A common reaction is to dismiss them as too improbable or too far in the future to be worthy of concern.[74] Fearon asks how we can 'constructively define something [superintelligence] that, by hypothesis, we cannot currently conceive of', which he seems to use as a justification for discounting the threat.[75] But that a threat is *hard to conceive of* need not mean it is *improbable*.[76] Viruses were hard to conceive of before the advent of modern medicine, yet biological warfare has turned out to be a real risk indeed. With a plethora of field-leading experts—many independent of industry—signing the statement 'Mitigating the risk of extinction from AI should be a global priority alongside other societal-scale risks such as pandemics and nuclear war', their warnings should be taken seriously.[77] Given the potential cost if AGI or ASI prove existentially dangerous, it would be foolish to ignore these concerns, even if one thinks they are probably unfounded.[78] Gambling on the hope that they will never materialize is, Stuart Russell observes, 'like saying we are driving towards the cliff but we're bound to run out of gas before we get there. And that doesn't seem like a good way to manage the affairs of the human race'.[79]

---

[70] Bales, 'AI takeover'; Grace et al., 'Thousands of AI authors', p. 41; Dan Milmo and Kiran Stacey, 'Experts disagree over threat posed but artificial intelligence cannot be ignored', *Guardian,* September 25, 2023, https://www.theguardian.com/technology/2023/sep/25/experts-disagree-over-threat-posed-but-artificial-intelligence-cannot-be-ignored-ai.
[71] Baum, Barrett and Yampolskiy, 'Modeling', p. 422; Dung and Hellrigel-Holderbaum, 'Against racing', p. 22; Simon Friederich, 'Symbiosis, not alignment, as the goal for liberal democracies in the transition to artificial general intelligence', *AI & Ethics,* 4: 2, 2024, p. 318, https://doi.org/10.1007/s43681-023-00268-7.
[72] Matthew Rendall, 'Macrosecuritisation failure and technological lock-in: lessons from the history of the bomb', *European Journal of International Relations,* 32: 1, 2026, pp. 59-83, https://doi.org/10.1177/13540661251398.
[73] Bostrom *Superintelligence,* pp. 24, 322; Carlsmith, 'Existential risk', pp. 386-87; cf. Yudkowsky and Soares, *If anyone builds it.*
[74] For a review and critique, see Yampolskiy, "AI risk skepticism', pp. 225-48.
[75] Fearon, 'Nuclear revolution', pp. 22-23.
[76] Olle Häggström, *Here be dragons: science, technology and the future of humanity* (Oxford: Oxford University Press, 2016), p. 244.
[77] Center for AI Safety, 'Statement on AI risk', 2023, https://www.safe.ai/statement-on-ai-risk; Leonard Dung, 'Evaluating approaches for reducing catastrophic risks from AI', *AI and Ethics,* 5: 2, 2025, p. 1178, https://doi.org/10.1007/s43681-024-00475-w; Gloor, 'Altruists'; Ng, 'Global extinction', p. 260, https://doi.org/10.1111/1758-5899.12647; Yampolskiy, 'AI risk skepticism', p. 233.

*AI racing and the loss of control*

Unfortunately, developing AGI without regard for its safety is faster and easier than developing AGI that is safe.[80] A race with all parties placing enormous value on attaining ASI first would likely reduce the time the world has to work on the alignment problem and divert both time and resources from safety research.[81] It would also make it much harder for the parties to the race—presumably the United States and China—to cooperate on essential AI safety research.[82] Furthermore, AI-racing is likely to promote secrecy, making it harder to have a robust public debate about the risks ASI development involves.[83] Perhaps worst of all, market competition or an arms race could encourage programmers to loosen restraints.[84] A 2020 study found 72 research programs on artificial general intelligence underway in 37 countries, several conducted by actors who dismissed safety concerns.[85] Some can be expected to gamble on ASI without prioritizing its alignment, given the huge potential benefits. Those prepared to throw safety out the window may get to ASI first.[86]

---

[78] Baum, 'Takeover catastrophe', p. 753; Jon Elster, 'Risk, uncertainty and nuclear power', *Social Science Information,* 18: 3, 1979, p. 383, https://doi.org/10.1177/053901847901800З; Stephen F. Haller, *Apocalypse soon? Wagering on warnings of global catastrophe* (Montreal: McGill-Queen's University Press, 2002).
[79] Quoted in Maureen Dowd, 'Elon Musk's billion-dollar crusade to stop the A.I. apocalypse', *Vanity Fair,* 26 March 2017, https://www.vanityfair.com/news/2017/03/elon-musk-billion-dollar-crusade-to-stop-ai-space-x.
[80] Stephen McAleese, 'How do AI timelines affect existential risk?', unpublished paper, August 2022, pp. 5-6, https://doi.org/10.48550/arXiv.2209.05459.
[81] Dung and Hellrigel-Holderbaum, 'Against racing', pp. 13-14.
[82] Ó hÉigeartaigh, Lu, Snell and Hashim, 'When the chips are down', p. 21.
[83] Dung and Hellrigel-Holderbaum, 'Against racing', p. 12; Ó hÉigeartaigh, Lu, Snell and Hashim, 'When the chips are down', p. 21.
[84] Stuart Armstrong, Nick Bostrom and Carl Shulman, 'Racing to the precipice: a model of artificial intelligence development', *AI & Society,* 31: 2, 2016, pp. 201-6, https://doi.org/10.1007/s00146-015-0590-y; Brundage, 'Unbridled AI competition', pp. 11-12; Hendrycks, 'Natural selection', pp. 30-31; Katzke and Futerman, 'Manhattan trap', pp. 15-16.
[85] Émile P. Torres, *Human extinction: a history of the science and ethics of annihilation* (New York: Routledge, 2023), pp. 184-85, citing McKenna Fitzgerald, Aaron Boddy and Seth D. Baum, *2020 survey of artificial general intelligence projects for ethics, risk and policy* (New York: Global Catastrophic Risk Institute, 2020).

A key issue is whether the winner of the race could capitalize on a technological lead to achieve a decisive advantage.[87] Worries of this kind could drive otherwise responsible actors to cut corners.[88]

> 'In the Cold War', observe Dan Hendrycks, Eric Schmidt and Aleksandr Wang,
>
> the phrase "Better dead than Red" implied that losing to an adversary was seen as worse than risking nuclear war. In a future AI race, similar reasoning could push officials to tolerate a double-digit risk of losing control if the alternative—lagging behind a rival—seems unacceptable. If the choice is stark—risk omnicide or lose—some might take that gamble. Carried out by multiple competing powers, this amounts to global Russian roulette and drives humanity toward an alarming probability of annihilation.[89]

While 'race to the bottom' dynamics can sometimes be controlled by regulation, this is challenging outside the bounds of the sovereign state.[90] Already we see critics of AI regulation, including the Trump administration, citing the risk of losing ground to competitors.[91] Even those trying to act responsibly are likely to disagree about what is safe. Development may be subject to the 'unilateralist's curse', with the least cautious actor determining the outcome. If a single actor proceeds with malevolent intent and achieves a decisive advantage or simply fails to align ASI, perhaps due to a lack of caution, the result could be catastrophe.[92]

---

[86] Nick Bostrom, 'Strategic implications of openness in AI development', *Global Policy,* 8: 2, 2017, pp. 139-40, https://doi.org/10.1111/1758-5899.12403; Carlsmith, 'Existential risk', pp. 404-2; Gloor, 'Altruists'; Hendrycks, 'Natural selection', p. 5.

[87] Lisa Abraham, Joshua Kavner and Alvin Moon, *A prisoner's dilemma in the race to artificial general intelligence* (Santa Monica, CA: RAND Corporation, 2025), https://www.rand.org/content/dam/rand/pubs/research_reports/RRA4200/RRA4245-1/RAND_RRA4245-1.pdf; MacAskill and Moorehouse, 'Preparing', pp. 33-36.

[88] Carlsmith, 'Existential risk', p. 401 n. 74; de Neufville and Baum, 'Collective action', p. 4.

[89] Hendrycks, Schmidt and Wang, *Superintelligence strategy,* p. 12; see also Katzke and Futerman, 'Manhattan trap', pp. 19-20; Shulman, 'Omuhundro's "Basic AI drives"', p. 3.

[90] Cf. Carlsmith, 'Is power-seeking AI an existential risk?', p. 36; Russell, *Human compatible,* pp. 135-36, 155-56.

[91] Chandelis Duster, 'Trump downplays calls for AI slowdown', *NPR,* 14 September 2026, https://www.npr.org/2026/09/13/nx-s1-5968078/trump-mike-johnson-ai-slowdown; Executive Office of the President, *Winning the race: America's AI action plan* (Washington: Executive Office of the President of the United States, 2025), p. 1, https://zw.usembassy.gov/winning-the-race-americas-ai-action-plan/; David E. Sanger and Dustin Volz, 'As fears of A.I. catastrophe magnify, Washington stirs, but mostly slumbers', *New York Times,* 13 September 2026, https://www.nytimes.com/2026/09/13/us/politics/ai-catastrophe-fears-washington.html.

Some have construed the risk of misalignment as an argument *for* racing. Writing in June 2024, Leopold Aschenbrenner recommended that the United States compete fiercely to stay ahead, in part because 'a healthy lead gives us room to maneuver: the ability to "cash in" parts of the lead, if necessary, to get safety right, for example by devoting extra work to alignment during the intelligence explosion'. Negotiating from this position of strength, Washington could then oversee the construction of an international ASI safety regime.[93] It is not clear just how this was to work. One of Aschenbrenner's fears was that superintelligence could enable enduring totalitarian rule. Given the enormous capabilities superintelligence would afford for coercion and manipulation, this was a reasonable fear.[94] Yet an American monopoly could also empower *internal* dictators.[95] Moreover, Aschenbrenner seemed to assume that the United States could achieve superintelligence without China launching a preventive war, then either intimidate Beijing into accepting its preferred safety regime or disarm and overthrow the Chinese regime at an acceptable cost.[96] This could be a tall order. Aschenbrenner also assumed that the United States would regard a safety regime as an important goal. Yet while the current Trump administration has announced its intention of pursuing 'global AI dominance', it has shown little concern for whether ASI would be properly aligned.[97] If ASI proves misaligned, the result could be human extinction.

---

[92] Daniel Dewey, 'Long-term strategies for ending existential risk from fast takeoff', in Vincent C. Müller (ed.), *Risks of artificial intelligence* (Boca Raton, FL: CRC Press, 2016), pp. 246-50; Leonard Dung, 'The argument for near-term human disempowerment through AI', *AI & Society,* 40: 3, 2025, pp. 1195-1208, https://doi.org/10.1007/s00146-024-01930-2; Growiec, 'Existential risk', p. 1694; Hendrycks, Schmidt and Wang, *Superintelligence strategy,* p. 10.
[93] Aschenbrenner, 'Free world'; see also Dario Amodei, 'Machines of loving grace: how AI could transform the world for the better', *Dario Amodei* (blog), October 2024, https://www.darioamodei.com/essay/machines-of-loving-grace; Samuel Hammond, 'We need a Manhattan project for AI safety', *Politico,* 8 May 2023, https://www.politico.com/news/magazine/2023/05/08/manhattan-project-for-ai-safety-00095779; MacAskill and Moorehouse, 'Preparing', p. 51.
[94] See William MacAskill, *What we owe the future* (New York: Basic Books, 2022), chapter 4; MacAskill and Moorehouse, 'Preparing', pp. 33-36.
[95] Katzke and Futerman, 'Manhattan trap', pp. 16-18.
[96] Cf. Naci Cankaya and Jakub Krys, 'Hawkish nationalism vs. international ai power and benefit sharing', *The Datacenter Lie Detector* (blog), 17 October 2024, https://nacicankaya.substack.com/p/hawkish-nationalism-vs-international; Hendrycks, Schmidt and Wang, *Superintelligence strategy,* pp. 12-14; Katzke and Futerman, 'Manhattan trap', pp. 10-13.
[97] Radhika Rajkumar, 'The head of US AI safety has stepped down. What now?' *ZDNET,* February 19, 2025, https://www.zdnet.com/article/the-head-of-us-ai-safety-has-stepped-down-what-now/; Sanger and

**Averting an ASI race**

The foregoing considerations make clear that China and the United States both have a vital interest in averting an ASI race. In recent years a narrative has taken hold in the United States that the two countries are already racing. It is far from clear that that is the case. We do not yet see the level of secrecy and technology controls that characterized the nuclear arms race.[98] Nor is China obviously talking or behaving—at any rate in public—as if is trying to beat the United States to ASI.[99] The 'narrative that China is locked in a race with the USA to reach artificial superintelligence, and translate this to lasting geopolitical dominance', Séan Ó hÉigeartaigh concludes, 'is almost entirely constructed and promoted by US-associated actors'. Some probably believe their own narratives, whereas others may be deliberately inflating this threat posed by China. Much of the lobbying comes from the tech industry,[100] resembling past cases in which groups with privileged access to information and a parochial interest in competitive policies exaggerated foreign threats.[101] Repeated loosening of the US export controls on highly capable

---

Volz, 'Fears of A.I. catastrophe'. In early 2025, the so-called Department of Government Efficiency targeted all probationary employees in the US government for dismissal. This meant dismissing the majority of federal employees working on artificial intelligence, including AI safety. Kelsey Piper, 'Trump wants America to shine on AI. DOGE Is blundering it', *Vox,* February 22, 2025, https://www.vox.com/future-perfect/400574/the-trump-administration-doesnt-know-what-it-wants-on-ai.

[98] Dung and Hellrigel-Holderbaum, 'Against racing', p. 19.

[99] ChinaTalk, 'Is China racing to AGI?' *ChinaTalk* (blog), April 2, 2025, https://www.chinatalk.media/p/is-china-agi-pilled; Elizabeth Gibney, 'China's plan to lead the world on AI regulation', *Nature,* 648, 18/25 December 2025, p. 504, https://doi.org/10.1038/d41586-025-03902-y; Mohammed Hosseini and Laura Caroli, 'Rather than framing AI Competition as a "race" with China, to drive innovation the US should promote greater local and global AI regulation', *LSE United States Politics and Policy* (blog), 2 April 2026, https://blogs.lse.ac.uk/usappblog/2026/04/02/rather-than-framing-ai-competition-as-a-race-with-china-to-drive-innovation-the-us-should-promote-greater-local-and-global-ai-regulation/; Triolo, 'Costly illusion'; cf. Jared Perlo, 'China is starting to talk about AI superintelligence, and some in the U.S. are taking notice', *NBC News,* 4 October 2025, https://www.nbcnews.com/tech/tech-news/china-starting-talk-ai-superintelligence-us-are-taking-notice-rcna234675.

[100] Connor Leahy, Gabriel Alfour, Chris Scammell, Andrea Miotti and Adam Shimi, *The Compendium, V1.3.1,* unpublished paper, 9 December 2024, chapter 6, https://www.thecompendium.ai/summary; Garrison Lovely, 'China hawks are manufacturing an AI arms race', *LessWrong* (blog), 20 November 2024, https://www.lesswrong.com/posts/KPBPc7RayDPxqxdqY/china-hawks-are-manufacturing-an-ai-arms-race; Ó hÉigeartaigh, 'Most dangerous fiction', quoted passage at p. 63; Ó hÉigeartaigh, Liu, Snell and Hashim, 'When the chips are down'; Grace Werner, 'The AI arms race isn't inevitable', *Palladium* (blog), 23 August 2024, https://www.palladiummag.com/2024/08/23/the-ai-arms-race-isnt-inevitable/.

[101] Jack Snyder, *Myths of empire: domestic politics and international ambition* (Ithaca, NY: Cornell University Press, 1991); cf. Elisabeth Siegel, Chloe Chadwick and Fergus Linley-Mota, 'Private sector firms are telling the stories—and calling the shots—about AI', *Oxford Political Review,* 13 January 2025,

computer chips, for example allowing Nvidia's H200 chips to be shipped to China under certain conditions, suggests that the US government also has not consistently adopted a racing posture.[102] But even if China is not racing at present, it may feel compelled to do so if the AI race narrative takes hold.[103]

Analysts have proposed a wide range of measures to mitigate the risk of a race, many with antecedents in proposals to control nuclear weapons.[104] Some have called for international restrictions on ASI development, and/or a worldwide moratorium on frontier AI development.[105] A key issue is whether such an agreement would be verifiable, and if so, whether states would accept the level of monitoring required to do so.[106] But even if so, in the long run, algorithmic improvements are likely to allow actors to develop ASI using widely accessible hardware.[107]

Rather than trying to restrict AI's capabilities, it may be more feasible to channel its development. A number of writers have recommended developing forms that give an advantage to defenders.[108] Whether and how the ASI offense-defence balance can be shifted in this way requires

---

https://oxfordpoliticalreview.com/2025/01/13/private-sector-firms-are-telling-the-stories-and-calling-the-shots-about-ai/.

[102] Zvi, 'Selling H200s to China is unwise and unpopular', *LessWrong* (blog), 9 December 2025, https://www.lesswrong.com/posts/kmEpWTjWeFyqv4tb5/selling-h200s-to-china-is-unwise-and-unpopular.

[103] Werner, 'AI arms race'.

[104] For overviews, see de Neufville and Baum, 'Collective action'; Nicholas Emery-Xu, Richard Jordan and Robert Trager, 'International governance of advancing artificial intelligence', *AI & Society,* 40: 4, 2025, pp. 3019-3044, https://doi.org/10.1007/s00146-024-02050-7, Sammy Martin, Justin Bullock and Corin Katzke, 'Analysis of global AI governance strategies', *LessWrong* (blog), 4 December 2024, https://www.lesswrong.com/posts/6nNwMbdRXZDuNd4Gx/analysis-of-global-ai-governance-strategies.

[105] Aguire, 'Close the gates'; Future of Life Institute, 'Pause giant AI experiments: An open letter', 22 March 2023, https://futureoflife.org/open-letter/pause-giant-ai-experiments/; Thomas Larsen et al., 'AI 2040: plan A', AI Futures Project, July 2026, https://ai-2040.com/; William MacAskill, *A global convention to govern the intelligence explosion* (Oxford: Forethought Research, 2026), https://www.forethought.org/research/a-global-convention-to-govern-the-intelligence-explosion; Andrea Miotti, Tolga Bilge, Dave Kasten and James Newport, 'A narrow path: how to secure our future', unpublished paper, April 2024, https://www.narrowpath.co/; Edward Roussel et al., 'Are we doomed to an AI race? Why self-interest could drive countries towards a moratorium on superintelligence', unpublished paper, 8 May 2026, https://doi.org/10.48550/arXiv.2605.01297; cf. Dung, 'Evaluating approaches', pp. 1183-85.

[106] Emery-Xu, Jordan and Trager, 'International governance', pp. 3034-38; Katzke and Futerman, 'Manhattan trap', pp. 22-26.

[107] Dung and Hellrigel-Holderbaum, 'Against racing', pp. 21-22; Helen Toner, 'Nonproliferation Is the wrong approach to AI misuse', *Rising Tide* (blog), 5 April 2025, https://helentoner.substack.com/p/nonproliferation-is-the-wrong-approach.

[108] For discussion, see Joe Carlsmith, 'AI for AI safety', *Joe Carlsmith* (blog), 14 March 2025, https://joecarlsmith.com/2025/03/14/ai-for-ai-safety/; Eric Drexler, 'Coercive cooperation: forcing win-

more research.[109] Others have proposed regulating ASI through an international agency,[110] or concentrating advanced AI research in an international monopoly along the lines of the Acheson-Lilienthal and Baruch Plans for atomic energy.[111] These proposals face some of the same obstacles that have long bedevilled proposals for the international control of atomic energy.[112] Yet collaborative research on AI safety might succeed even if other initiatives fail. Collaboration would increase the chance of the first actor to deploy ASI having the knowledge needed to align it properly.[113] An important first step would be to bring Chinese experts into existing international forums on AI safety.[114]

Even unilateral research could help make all countries' practices safer. No state has an interest in its rivals losing control of ASI. This gives all an incentive to disseminate solutions to AI alignment, just as during the Cold War the United States shared information about permissive

---

win outcomes in an AI-driven transition', *AI Prospects: Toward Global Goal Alignment* (blog), May 22, 2025, https://aiprospects.substack.com/p/coercive-cooperation-forcing-win; MacAskill and Moorhouse, 'Preparing for the intelligence explosion', p. 33; Lizka Vaintrob and Owen Cotton-Barratt, 'Appendices to AI tools for existential security', Forethought Research, 14 March 2025, https://www.forethought.org/research/appendices-to-ai-tools-for-existential-security.

[109] Toby Ord notes the undertheorized state of the issue. *The precipice: existential risk and the future of humanity* (London: Bloomsbury, 2020), p. 208.

[110] Sam Altman, Greg Brockman and Ilya Sutskever, 'Governance of superintelligence', *OpenAI* (blog), 23 May 2023, https://openai.com/index/governance-of-superintelligence/; Tom Davidson, 'The danger of runaway AI', *Journal of Democracy,* 34: 4, 2023, pp. 137-39, https://doi.org/10.1353/jod.2023.a907694; McLean et al., 'Risk associated with artificial general intelligence', p. 657; Emery-Xu, Jordan and Trager, 'International governance', pp. 3039-3040.

[111] Miles Brundage, 'My recent lecture at Berkeley and a vision for a "CERN for AI"', *Miles's Substack* (blog), 11 December 2024, https://milesbrundage.substack.com/p/my-recent-lecture-at-berkeley-and; William MacAskill, *International AI projects and differential AI development* (Forethought Research, 2026), https://www.forethought.org/research/international-ai-projects-and-differential-ai-development; Miotti, Bilge, Kasten and Newport, 'A narrow path'; Miotti and Wasil, 'Taking control'; cf. Elliot Jones, 'A "CERN for AI"—what might an international AI research organization address?" in Alex Krasodomski (ed.), *Artificial intelligence and the challenge for global governance: nine essays on achieving responsible AI* (London: Chatham House, 2024), pp. 10-17; Emery-Xu, Jordan and Trager, 'International governance', pp. 3038-3039; Martin, Bullock and Katzke, 'Global AI governance strategies'.

[112] Benjamin Boudreaux, Gregory Smith, Edward Geist and Leah Dion, *Insights from nuclear history for AI governance* (Santa Monica, CA: RAND Corporation, 2025), https://www.rand.org/pubs/perspectives/PEA3652-1.html.; Dewey, 'Long-term strategies', p. 254; cf. Emery-Xu, Jordan and Trager, 'International governance', p. 3039; Waqar Zaidi and Allan Dafoe, *International control of powerful technology: lessons from the Baruch Plan for nuclear weapons* (Oxford: Future of Humanity Institute, 2021), pp. 4-5.

[113] Grace, 'Friendly AI'; Turchin, Denkenberger and Green, 'Global solutions', pp. 11-12, 17; cf. Martin, Bullock and Katzke, 'Global AI governance strategies'.

[114] Triolo, 'Costly illusion'.

action links—technology hindering the unauthorised use of nuclear weapons—to the USSR.[115] Sharing such information would signal cooperative intentions and the desire to avoid an AI race.[116] The flipside of the 'unilateralist's curse' is if that a single state finds ways to handle the alignment problem and shares them, other states are likely to adopt them.[117]

**Conclusion**

A narrative has taken hold that a race to superintelligence is inevitable. 'AI is today and will continue to be at the forefront of the fiercest Thucydidean rivalry history has ever seen', Graham Allison advises. Fortunately, 'the United States can compete—and win'.[118] But so far, only Washington appears to be promulgating the narrative of an ASI race. Should it goad Beijing into a contest, it would run three serious risks. The first two are familiar from past arms races: the United States could lose the race or provoke conflict that might turn nuclear. The third could be far worse: ASI escaping human control. Unless the superpowers work together to address this threat, we risk a scenario where AI itself wins the race, and all humanity loses.[119]

[115] Simon Friederich and Leonard Dung, 'Against the Manhattan Project framing of AI alignment', *Mind & Language,* 41: 1, 2026, p. 123, https://doi.org/10.1111/mila.12548 Jonas B. Sandbrink, Hamish Hobbs, Jacob L. Swett, Allan Dafoe and Anders Sandberg, 'Risk-sensitive innovation: leveraging interactions between technologies to navigate technology risks', *Science and Public Policy,* 51: 6, 2024, p. 1034, https://doi.org/10.1093/scipol/scae043; Matt Sheehan, 'China's views on AI safety are changing—quickly', *Carnegie Endowment for International Peace* (blog), 27 August 2024, https://carnegieendowment.org/research/2024/08/china-artificial-intelligence-ai-safety-regulation?lang=en.

[116] Brundage and Werner, 'America First meets safety first'.

[117] Anders Sandberg, 'Differential technology development: speeding up to slow down disaster?' unpublished paper, https://ora.ox.ac.uk/objects/uuid:a244c415-4824-49f1-801e-25f118be9cbe/files/skw52j979w; Turchin, Denkenberger and Green, 'Global solutions', pp. 11, 17; Vaintrob and Cotton-Barratt, 'Appendices to AI tools', p. 10.

[118] Allison, 'Clash of AI superpowers', pp. 22-23; *idem*, *The National Insecurity of AI* (Cambridge, MA: Belfer Center for Science and International Affairs, 2024), p. 2, https://www.belfercenter.org/research-analysis/national-insecurity-ai.

[119] Grace, 'AI is not an arms race'; Ó hÉigeartaigh, 'Most dangerous fiction', p. 64.